# Large positive magnetocaloric effect in a perovskite manganite

R. Mahendiran, B. G. Ueland and P. Schiffer

Department of Physics and Materials Research Institute, 104 Davey Laboratory,

Pennsylvania State University, University Park, PA  16802

A. Maignan, C. Martin, M. Hervieu, and B. Raveau

Laboratoire CRISMAT, ISMRA, Université de Caen, 6 Boulevard du Marèchal Juin,

Caen Cedex, Caen-14050, France

M. R. Ibarra and L. Morellon

Departamento de Física de la Materia Condensada, ICMA-CSIC, Universidad de

Zaragoza, Zaragoza-50009, Spain

We investigate the magnetothermodynamic properties of the perovskite manganite $Pr_{0.46}Sr_{0.54}MnO_3$. Our data imply that the paramagnetic to antiferromagnetic transition at $T_N$ = 210 K is first order with an abrupt decrease of volume (~ 0.14 % in zero magnetic field) and is accompanied by a sharp anomaly in specific heat.  Upon application of a sufficiently large magnetic field, the antiferromagnetic phase transforms into a ferromagnetic phase with a sharp increase in sample volume. This field induced transition results in a large positive magnetocaloric effect just  below $T_N$ ($\Delta S_m$ = 10.4 $Jkg^{-1}K^{-1}$ at  H = 5.5 T), which is associated with the increasing stability of the antiferromagnetic state with decreasing temperature.

Magnetic materials showing a large magnetocaloric effect (MCE), have attracted considerable recent attention for their potential application in magnetic refrigeration technology.[1,2] An especially large magnetic entropy change is expected near a first order magnetic transition accompanying a structural transition, and much interest has thus been focused on the MCE near the first-order magnetostructural phase transitions driven by a magnetic field, as in $Gd_5(Ge_{1-x}Si_x)_4$,[3] $MnAs$,[4] $MnFe_{0.45}As_{0.55}$,[5] $LaF_{11.6}Si_{1.4}C_x$,[6] and Ni-Ga-Mn alloys,[7] all of which also undergo temperature-driven first order magnetic transitions. Additionally, the colossal magnetoresistive manganese oxides ($R_{1-x}A_xMnO_3$, where R and A are trivalent rare earth ions and divalent alkaline earth ions respectively) which undergo temperature driven paramagnetic to ferromagnetic transitions, show relatively large "negative" MCE, in which the isothermal magnetic entropy change ($\Delta S_m = S(H, T) - S(0,T)$) is negative.[8] By contrast, when an external field is applied to an antiferromagnet, the magnetic entropy is expected to increase, but MCE studies on antiferromagnetic manganites are relatively scarce. A large positive $\Delta S_m$ (of order 7 $Jkg^{-1}K^{-1}$ at 1 T) can be observed in MCE data taken on $Pr_{0.5}Sr_{0.5}MnO_3$,[9] and a large MCE has been reported in charge ordered $Nd_{0.5}Sr_{0.5}MnO_3$ but without discussion of the sign of $\Delta S_m$.[10]

In this paper we study the magnetothermodynamics of the antiferromagnetic manganite, $Pr_{0.46}Sr_{0.54}MnO_3$. The paramagnetic (PM) to antiferromagnetic (AF) transition in zero magnetic field ($T_N$ = 210 K) is accompanied by a tetragonal to orthorhombic structural change[11] and orbital ordering.[12] Our data indicate that application of an external magnetic field drives the spins from the antiferromagnetic to a ferromagnetic state, as is common in the manganites. This transition results in a large structural change, and it is accompanied by a large positive MCE in the antiferromagnetic state which is comparable in magnitude to the largest negative MCE reported in the manganite compounds.

We have measured the thermal expansion, magnetization, resistivity, specific heat and magnetostriction and MCE of polycrystalline $Pr_{0.46}Sr_{0.54}MnO_3$. The sample was prepared by the standard solid state route and found to be single phase (tetragonal, I4/mcm) at 300 K by X-ray diffraction. The temperature dependence of the linear thermal expansion ($\Delta L/L$) in zero field was measured with a strain gauge between 300 K and 10 K. Magnetostriction isotherms in a pulsed magnetic field up to H = 14.2 T were measured with a strain gauge with measuring direction parallel and perpendicular to the magnetic field direction. Based on the measured parallel ($\lambda_{par}$) and perpendicular ($\lambda_{per}$) magnetostrictions, the volume magnetostriction was calculated from the relation w = $\lambda_{par}+2\lambda_{per}$. Magnetization was measured with a Quantum Design SQUID magnetometer, and resistivity was measured with a standard 4-probe technique. The specific heat in zero magnetic field was measured by the relaxation method using a Quantum Design Physical Property Measuring System.

Figure 1(a) shows the temperature dependence of the volume expansion ($\Delta V/V = 3*\Delta L/L$) of $Pr_{0.46}Sr_{0.54}MnO_3$. We find that $\Delta V/V$ decreases monotonically with temperature, with an abrupt change (~0.14 %) near $T_N$ = 210 K. The resistivity ($\rho$) also shows an abrupt jump near $T_N$ (see figure 1(b)). Although $\rho(T)$ increases below $T_N$, its value is much smaller (= 60 m$\Omega$ cm at 5 K) compared with other antiferromagnetic manganites. This is a signature of the A-type AF state in which electrical conduction is dominated by the charge transport in two dimensional ferromagnetic metallic planes which are coupled antiferromagnetically along the c-axis.[11-12] Above $T_N$, $\rho(T, 0)$ shows a broad maximum around T = 225 K possibly due to the onset of short range ferromagnetic order (the established phase diagram[11] of $Pr_{1-x}Sr_xMnO_3$ suggests that $T_C$ and $T_N$ may be close to each other for x = 0.54). When the sample is cooled in an external magnetic field ( H ≤ 7 T) from 300 K, this maximum appears to be shifted above 300 K, and $T_N$ is shifted down as expected due to the larger Zeeman energy associated with antiferromagnetic order. Hysteresis between cooling and warming in both ρ(T) and ΔV/V, implies that the AF transition at $T_N$ is first order in character. The inset of figure 1(b) shows the temperature dependence of the magnetization, which displays a sharp drop near $T_N$.

The paramagnetic-antiferromagnetic transition is also accompanied by a sharp peak in the specific heat (C) which is shown in figure 2. The inset shows the excess specific heat (ΔC/T) due to the phase transition, subtracting the background by fitting with the Einstein model, $C = \sum_i a_i \left[ x_i^2 e^{x_i} / (e^{x_i} - 1)^2 \right]$ where i = 1, 2, 3 are three optical vibrational modes and $x_i = h\nu_i/k_BT$ where $h\nu_i/k_B$ = 455 K, 168 K and 50 K (the fit is shown by the dotted line). The entropy loss in this transition (obtained by integrating the area under ΔC/T curve) is ~2.5 J mole$^{-1}$K$^{-1}$. As expected, this value is smaller than the total spin entropy $(0.46Rln(2S_{Mn}^{3+}+ 1)+0.56Rln(2S_{Mn}^{4+}+1))$ = 12.6 Jmole$^{-1}$K$^{-1}$, where R is the gas constant, $S_{Mn}^{3+}$ = 2, and $S_{Mn}^{4+}$ = 3/2.

Figure 3 (a) shows the field dependence of the magnetization M(H) data in the paramagnetic state below T = 270 K. The value of M increases with decreasing T, and, as T approaches $T_N$, M(H) increases more rapidly at low fields. This is possibly due to reorientation of short range ferromagnetic clusters with magnetic field. (Note that zero field ρ(T) showed a weak maximum above $T_N$, possibly due to double exchange within these clusters). The M(H) data below 210 K (figure 3(b)) are qualitatively different from the high temperature data. At T < 210 K, M(H) increases linearly with H until a critical field, $H_C$, at which there is a sharp rise to near saturation associated with the nucleation and rapid growth of the ferromagnetic domains in the antiferromagnetic matrix. The reverse transition (FM to AFM) is hysteretic which indicates that this field-induced transition is first order. The critical field for the jump in M(H) increases from H = 0.52 T at T = 210 K to more than 7 T at T = 187.5 K. The width of the hysteresis loop ($\approx$ 0.45 T) in the same temperature range is approximately constant. This is in contrast to the observed widening of the hysteresis loop with decreasing temperature in $Pr_{0.5}Sr_{0.5}MnO_3$ and $Nd_{0.5}Sr_{0.5}MnO_3$.[13]

The field-induced AF-FM transition is also manifested in the structure of this material. Figure 4 shows the field dependence of the volume magnetostriction at selected temperature. While the spontaneous volume decreases sharply upon cooling through $T_N$ (see figure 1(a)), the data in figure 4 show an abrupt volume *increase* associated with the field-induced AFM-FM transition. The volume magnetostriction is negligible in the paramagnetic state above 210 K (not shown in figure 4), but below $T_N$ it increases with lowering temperature to ~ 0.25 % at T = 165 K for our maximum field (H = 14.2 T). A similarly large volume expansion under a magnetic field has been previously reported in $Nd_{0.5}Sr_{0.5}MnO_3$ and $La_{0.5}Ca_{0.5}MnO_3$, associated with a structural transition from a low volume monoclinic (AF) to a high volume orthorhombic (FM) phase.[14] We therefore hypothesize that, in $Pr_{0.46}Sr_{0.54}MnO_3$, the external magnetic field induces a structural transition from orthorhombic (low volume) to tetragonal (high volume) symmetry. Since the A-type antiferromagnetism and tetragonal-orthorhombic transition in this compound in zero field are driven by ordering of $d_{x^2-y^2}$ orbitals,[11] the field-induced structural transition can also be considered as an orbital order-disorder transition.

Since a large MCE is anticipated near a first order field-induced magnetic transition, we calculate the magnetic entropy change, $\Delta S_m$, between zero field and a maximum field ($H_0$) using the thermodynamic relation:

$$\Delta S_m(T,H_0) = S_m(T, H_0) - S_m(T, 0) = \frac{1}{\Delta T}\int_0^{H_0} H\left[M(T+\Delta T,H)-M(T,H)\right]dH$$

where $\Delta T$ is the temperature increment between measured magnetization isotherms ($\Delta T$ = 5 K for our data).[1] The main panel of figure 5 shows the temperature dependence of the $\Delta S_m$ at H = 3 T and H = 7 T. We observe a positive peak in $\Delta S_m(T)$ around T = 210 K (205 K) for H = 3 T (7 T). The inset of figure 5 shows the calculated isothermal magnetic entropy change as a function of the magnetic field at different temperatures. For T ≥ 215 K, $\Delta S_m$ is negative and increases in magnitude nearly linearly with H. For T ≤ 215 K, however, $\Delta S_m$ is positive and shows a broad maximum and then decreases with decreasing temperatures. The magnitude of the maximum entropy change increases with decreasing temperature from 3 Jkg$^{-1}$K$^{-1}$ at T = 215 K (H = 2.25 T) to $\Delta S_{max} \approx 10.4$ Jkg$^{-1}$K$^{-1}$ at T = 205 K (H = 5.5 T). At lower temperatures, the maximum entropy change is above the field range of our measurements.

The negative $\Delta S_m$ above $T_N$ can be understood as a result of the field induced reorientation of ferromagnetic clusters which decreases the magnetic entropy. This is similar to the negative magnetocaloric effect which has been observed in many ferromagnetic and paramagnetic materials. The crossover to positive $\Delta S_m$ below $T_N$ is also expected in an antiferromagnetic state, and is consistent with data shown (but not discussed in detail) in the work of Chen et al. The large magnitude of the positive $\Delta S_m$, is quite surprising, but it can be attributed to the proximity of the field induced ferromagnetic transition which is strongly coupled to the structural transition in this material.

The observed maximum magnitude of $\Delta S_m = 10.4$ Jkg$^{-1}$K$^{-1}$ in H = 5.5 T is still lower than that of the giant negative magnetocaloric effect ($\Delta S_m = -19$ Jkg$^{-1}$K$^{-1}$ at H = 5 T) observed in GdSiGe alloy.[2] The value does compare favorably, however, with previous reports of large negative MCE in the manganites, such as the -7.9 Jkg$^{-1}$K$^{-1}$ at T = 244 K and H = 8 T previously reported in La$_{0.84}$Sr$_{0.16}$MnO$_3$. The maximum $\Delta S_m$ in our compound also occurs about 50 K higher than that observed in Pr$_{0.5}$Sr$_{0.5}$MnO$_3$ due to the higher $T_N$. The observation of a large positive magnetocaloric effect suggests an alternate method of magnetic refrigeration – through raising rather than lowering the magnetic field. Since both the Néel temperature and the magnetic field scale of the field induced transition can be altered either by A-site substitution (for example, (Pr$_{1-x}$La$_x$)$_{0.46}$Sr$_{0.54}$MnO$_3$) or by Mn site substitution (for example, Pr$_{0.46}$Sr$_{0.54}$Mn$_{1-x}$Cr$_x$O$_3$)[15], this effect has the possibility to expand the potential use of manganites in magnetocaloric refrigeration.

Acknowledgments: This work was supported by NSF grant DMR-101318. R.M. expresses thanks to MNERT (France) and MyCT (Spain) for partial financial assistance.

Fig. 1: Temperature dependence of the (a) fractional volume change ($\Delta V/V$) and (b) the resistivity in H = 0 and 7 T. The structural symmetry on either side of the Néel transition ($T_N$ = 210 K) was taken from Ref. 11. The inset of Fig. 1(b) shows the temperature dependence of the magnetization at three different magnetic fields ( H = 0.01 T, 1 T and 5 T) measured on cooling from T = 320 K.

Fig. 2: Temperature dependence of specific heat (C) while cooling in zero magnetic field from 300 K. The large peak in the specific heat reflects the strongly first order nature of the phase transition at $T_N$. The dashed line is a fit to the Einstein model to the phonon entropy. The inset shows the excess specific heat over temperature ($\Delta C/T$) after subtracting the Einstein contribution from the experimental curve.

Fig. 3: Field dependence of the magnetization at selected temperatures (a) above $T_N$ = 210 K and (b) below $T_N$. In figure 2(a) the temperature increment was 5 K and in figure 2(b) data at only few temperatures were shown for clarity although data were taken for each 2.5 K below 215 K.

Fig. 4: Field dependence of the volume magnetostriction at selected temperature. The positive volume magnetostriction implies a field induced structural transition from the low volume orthorhombic to the high temperature tetragonal phase. The volume magnetostriction above 215 K is negligible and hence not shown here.

Fig. 5: Temperature dependence of the magnetic entropy change ($\Delta S_m$) at H = 3 T and 7 T. Note that $\Delta S_m$ changes sign at $T_N$. The inset shows the field dependence of the magnetic entropy calculated from the M(H) data.

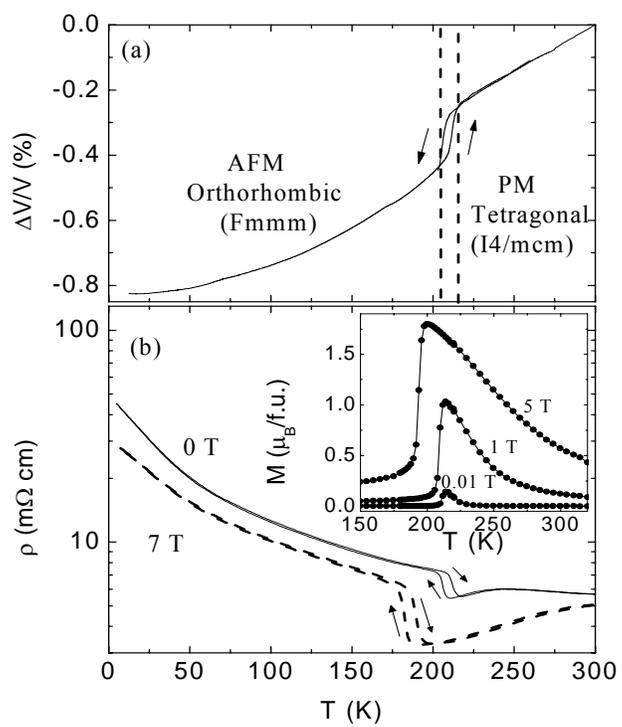

Fig. 1
R. Mahendiran et al.

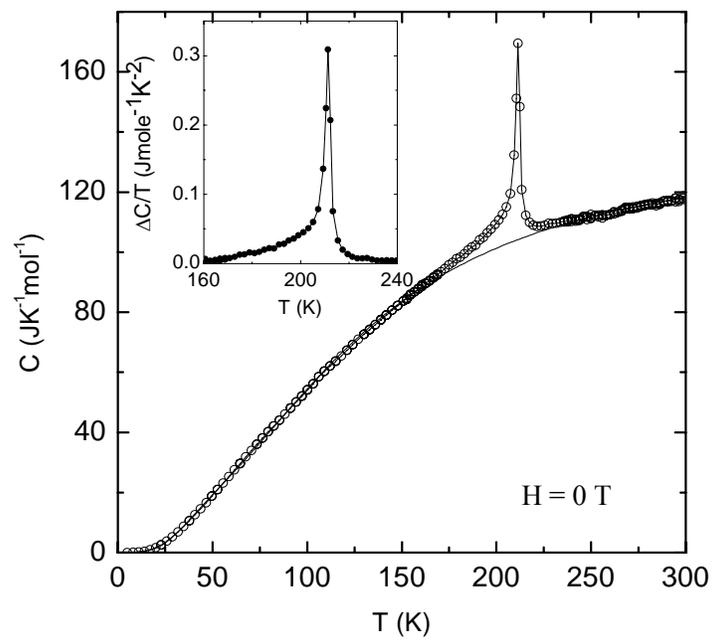

Fig. 2
R. Mahendiran et al.

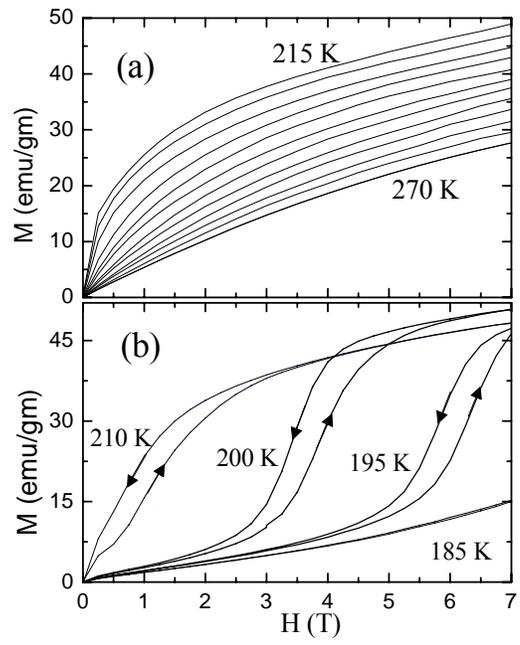

Fig. 3
R. Mahendiran et al.

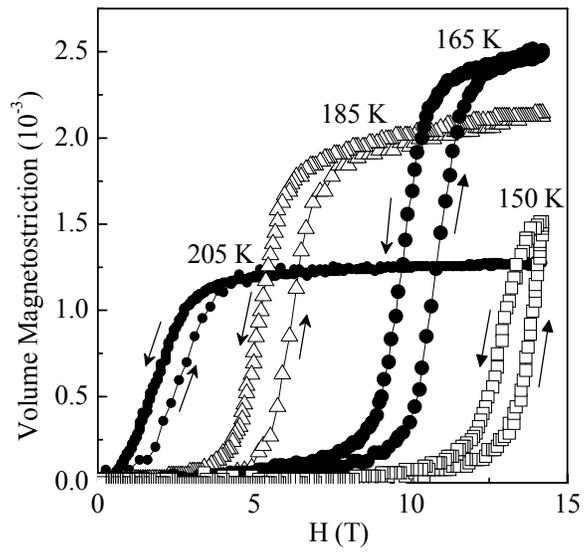

FIG. 4
R. Mahendiran et al.

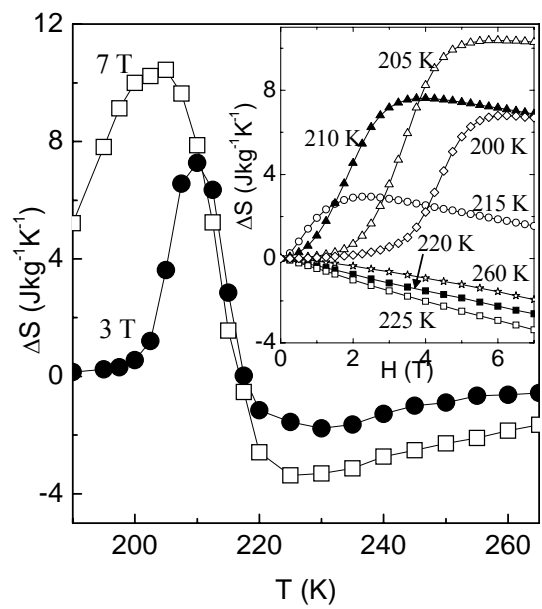

Fig. 5
R. Mahendiran et al.